**Atomic resolution imaging of the strength and 3D direction of atomic displacements in functional oxide thin films**

Ian MacLaren[1*], Aurys Silinga[1], Juri Barthel[2], Josee Kleibeuker[3,4], Judith L MacManus-Driscoll[3], Christopher S Allen[5,6], Angus I Kirkland[5,6]

[1]School of Physics and Astronomy, University of Glasgow, Glasgow G12 8QQ, UK
[2]Ernst Ruska-Centre (ER-C 2), Forschungszentrum Jülich GmbH, 52425 Jülich, Germany
[3]Department of Materials Science & Metallurgy, University of Cambridge, 27 Charles Babbage Road, Cambridge CB3 OFS, UK
[4]DEMCON high tech systems Eindhoven BV, Kanaaldijk 29, 5683 CR Best, The Netherlands
[5]electron Physical Science Imaging Centre, Diamond Light Source Ltd., OX11 0DE, UK
[6]Department of Materials, University of Oxford, Parks Road, Oxford OX1 3PH, UK

**ABSTRACT**. Atomic resolution imaging is key to understanding thin film growth and how a particular set of conditions influences properties. Whilst such imaging in the scanning transmission electron microscope (STEM) has had transformative impact in nanoscience, it forms projection images and provides no direct information about displacements perpendicular to the image plane. In this article we show that it is possible to make atomic resolution maps of the direction and magnitude of La displacements at ~30° to the imaging plane in a $La_2CoMnO_6$ thin film on (111) LSAT ($LaAlO_3$-$La(Sr,Ta)O_3$) using a 4-dimensional STEM (4DSTEM) methodology. This reveals that the La modulation lies preferentially in the interface plane, and is strongly suppressed close to the epitaxial interface, and further reveals how the modulation varies with distance from the interface with unit cell resolution. These details would be completely invisible to all prior techniques in electron microscopy and this sheds light on why this particular substrate in this orientation best promotes double perovskite cation ordering, and the consequent optimal magnetic ordering for this thin film system. The approach used herein of fitting atomic resolution 4DSTEM data to determine crystal parameters opens the door for a new era of atomic-resolution crystallography.

## I. INTRODUCTION.

Atomic resolution imaging in the scanning transmission electron microscope (STEM) has revolutionised science and technology, especially since the introduction of aberration correction[1,2], making sub-Ångström resolution easily realised. The ability to simply see two-dimensional projections of the atomic structure in thin specimens of crystalline solids has allowed unprecedented understanding of local features in nanostructures, such as interfaces, as well as of defects within materials and heterostructures [3-5]. In specific cases, more information about the 3-dimensional structure can be determined by looking at multiple examples of the same structure using discrete tomography [6-8], or from conventional tilt-tomography, although the latter is hard to perform at atomic resolution [9] at reasonable fluence levels. Recently, however, major advances have been made in STEM with the advent of fast-readout direct electron detectors for 4DSTEM [10,11], which offers possibilities for novel imaging modalities not possible with monolithic annular [12-14], circular [15] or segmented detectors [16,17]. We have previously shown that high angle electron diffraction into higher-order Laue zones produces an atomic resolution signal [18], which reveals information about the 3D order in a crystal. This current work goes far beyond that and shows that the atomic resolution diffraction data contains information about the exact 3D orientation of one unique axis of the unit cell and the magnitudes of atom movements along it; and that this can be mapped at atomic resolution by fitting a simple mathematical model. In short, we show that by measuring and fitting high angle scattering, 4DSTEM can perform atomic resolution imaging of 3D crystal ordering without any need for sample tilting. Such approaches will provide a major step forward in understanding the details of nanoscale heterostructures, including surface reconstructions and interface structures.

Characterising the local crystallographic structure of materials is key to understanding their behaviour and how the growth or fabrication leads to a given structure with specific properties. However, despite the ability of electron microscopy to image nanostructures with atomic resolution, this remains a two-dimensional projection and gaining access to information in the third dimension is very challenging in dimensionally constrained samples such as few-nm thin films or domain structures. Such considerations are especially pertinent when considering the optimisation of functional oxides and other functional materials, which are often grown as epitaxial systems with interfaces exerting a significant control over the resulting structure, chemical ordering and properties, particularly through strain transfer. However, we previously demonstrated that some information about the periodicity in the third dimension was available using a *HOLZ-STEM* (higher-order Laue zone - scanning transmission electron microscopy) approach[19], and that this is atomically resolved[18] (as also shown without 4DSTEM by Huang *et al.*[20]). This only recently became convenient to perform because of the advent of 4DSTEM (4-dimensional scanning transmission electron microscopy) in which a scan is performed over an area with a diffraction pattern recorded to a suitably fast detector at every scan point[10,11]. Nevertheless, more directional information than this about atomic displacements was absent until now.

*Contact author: ian.maclaren@glasgow.ac.uk

A common feature of functional oxides is octahedral coordination of the smaller cations and the coordinated tilting of these octahedra, often coupled with antiparallel displacements of the larger cations [21]. Whilst octahedral tilting is particularly common in perovskites and there exists a specific classification scheme for the tilting patterns thereof [22], similar effects are observed in a much wider range of functional oxides. This is particularly likely where the unit cell contains more than one similar layer and the cation is located away from a centrosymmetric position, as is seen, for example, in some tungsten bronzes[23]. This work particularly focuses on $La_2CoMnO_6$ (LCMO), which can order as a double perovskite with the B-site Co and Mn forming charge order as $Co^{2+}$ and $Mn^{4+}$ and ordering on two interpenetrating face-centred lattices in a monoclinic cell ($P2_1/n$) with $a$ = 5.52464 Å, $b$ = 5.48302 Å, $c$ = 7.7717 Å, $\beta$ = 89.898° (as refined by Bull *et al.*[24]). **Figure 1** shows a side view of a one-atom-thick slice of the structure of LCMO, (although with displacements exaggerated to make them clearer to the reader), showing such antiparallel displacements of alternate La atoms along the [111] direction (vertical in the figure), the displacements from equally spaced positions when viewed along a $\langle 111\rangle_M$ or $\langle 11\bar{1}\rangle_M$ direction being parallel to the **b** direction (actually $\mathbf{d} = \pm 0.0216[010]_M$, c.f. Supplemental Information section S1 for derivation). A key point is that rotating the crystal by 180° about the beam direction ($[111]_M$ in the figure) does not result in equivalent 3D structures, one showing an up-left, down-right pattern, and the other an up-right, down-left pattern. Despite this clear difference in 3D structure, projection images of any sort (such as annular dark field STEM) taken along $[111]_M$ from crystals represented by the two panes of Figure 1 would be identical. A 3D view of these coordinated atom movements and how they relate to the sample overall (e.g. at interfaces and strain fields) would be invaluable in better understanding the growth of complex oxides, especially in the presence of interfaces and/or multiple domains on the nanoscale. Either of these would make tilt tomography approaches impracticable. This would in turn inform the optimisation of the growth to produce atomic structures with the desired properties. The present work shows that using a 4DSTEM / HOLZ-STEM approach, it is possible to determine these out-of-plane displacements at atomic resolution and use those to map out the displacement patterns close to an epitaxial interface.

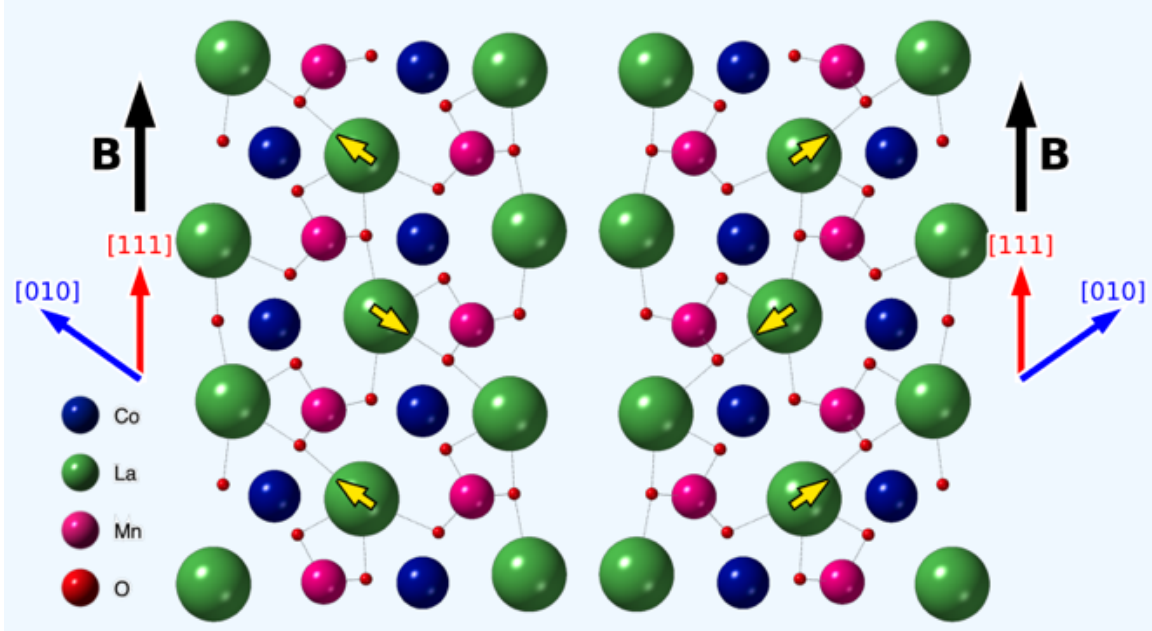

FIG 1. Schematic view of columns of atoms along the $[111]_M$ direction (vertical) in $La_2CoMnO_6$ in which there are alternate antiparallel La cation displacements along a direction parallel to $[010]_M$. Two possible orientations of the crystal are shown, containing the same vertical $[111]_M$ direction, but differing by a 180° rotation about this vertical direction). These two crystal orientations would give indistinguishable projection images when recorded with a beam direction parallel to $[111]_M$ (i.e. the 2D projection symmetry includes a 2-fold rotation)), however the 3D orientation of the **b** axis is switched. Just one plane of the crystal is shown for simplicity. La atom displacements are magnified by a factor of 3 to aid visualisation.

## II. METHODS

### A. Sample preparation and scanning transmission electron microscopy

The LCMO sample was grown by pulsed laser deposition on a (111) substrate of LSAT ($LaAlO_3$-$(La,Sr)TaO_3$) at a substrate temperature of 710°C using a repetition rate of 1Hz and a fluence of 1.5 $Jcm^{-2}$ under 0.15 mbar flowing $O_2$, followed by annealing at 700°C for 30 min under 500 mbar $O_2$, and finally slow cooling to room temperature. A sample was prepared for STEM using a standard FIB liftout procedure. Further details are given in Kleibeuker *et al.* [25].

Scanning transmission electron microscope datasets were acquired using a JEOL ARM300F at the ePSIC facility at Diamond Light Source operated at 200 kV with the diffraction patterns acquired using a 2x2 tiled Medipix detector providing 515x515 pixel images (Quantum Detectors Ltd., UK). The convergence angle of the probe was 20 mrad and the camera length was set to allow angles of up to about 97 mrad to reach the detector (centre to middle of edge), and a step size of ~0.5 Å was used to collect the dataset shown. The microscope defocus was tuned to give the best atomic-resolution HAADF imaging since this should give the best resolution of diffraction information in 4DSTEM and defocus can be considered to be close to 0 nm or very slightly underfocused (i.e. focused in top of the sample). No attempts were made to vary defocus or examine the effects of doing so experimentally, although simulations were performed on the effects thereof and these are in the Supplemental Information, Section S4. In summary, the main features in the parameter maps shown below are shown as long as good atomic-resolution HAADF images are produced.

### B. Data handling and initial processing (polar transform and fitting)

Conversion of the raw data to hdf5 was performed on site at Diamond Light Source, and all calculations were done on datasets where the diffraction dimension had been binned by a factor of 4 prior to further processing, since this both reduced memory requirements and increased processing speed, with no noticeable loss of any information in the results of the analysis. The centre of the FOLZ ring in the dataset was determined by overlaying a circle on the data and shifting this until it best matched the ring (as the centre-of-mass of the diffraction pattern, dominated by low angle scattering, can differ slightly from the centre of the FOLZ, if there is slight sample mistilt). This was then followed by performing the polar transform and testing if the Laue Zone ring is a straight line after transform.

Polar transform over entire 4DSTEM datasets was performed using home-built code (provided in the data deposit, alongside all other code pieces and raw data used in this work) based on the earlier implementation of Barthel in *emilys*, using the same maths, but optimised for fast calculation of an entire dataset. The relevant feature of the FOLZ ring was identified (the outer part of the split ring[18]) corresponding most strongly to the La-O columns and integrated to a 1D line plot of intensity as a function of azimuthal angle.

This was then fitted to Equation (1):

$$I(\phi) = \begin{array}{c} A_2 \cos^2(\phi - \phi_2) \\ +A_1 \cos(\phi - \phi_1) \\ +B \end{array} \quad (1)$$

(which is the fit function described in our previous , although some parameter renaming has been performed to aid clarity of understanding the meaning of each parameter from its name). $A_2$ is the amplitude of the 2-fold intensity modulation, $\phi_2$ is the peak angle of that 2-fold modulation (defined in the range 0-180°), $A_1$ is the amplitude of the 1-fold intensity modulation, $\phi_1$ is the angular shift of that 1-fold modulation (defined in the range 0-360°) and $B$ is the offset from zero. Fitting to the function was performed using *scipy.optimize.curve_fit* using a simple iterative function that fits each pixel in the real space scan in turn.

### C. A note on structures and notation

LSAT is pseudocubic (notwithstanding nanoscale chemical and likely structural variations), and directions and planes referenced to a primitive cubic perovskite cell have the subscript *C*, e.g. $(111)_C$.

LCMO, as noted above, is monoclinic with $a, b, c \approx \sqrt{2}a_C, \sqrt{2}a_C, 2a_C$. Directions and planes referenced to this cell have a subscript *M*, e.g. $[111]_M$. The A-site modulation is close to $[010]_M$. "**b**-axis" only ever refers to the *M* structure.

Some of the related structures discussed in the discussion have orthorhombic structures, which can either be expressed as *Pbnm* or *Pnma*, two alternate settings of space group 62. *Pbnm* is the easier of the two to relate to the monoclinic cell, as the axes come in exactly the same order with similar relative lengths. *Pnma* is also used, with the A-site modulation close to $[100]_{Pnma}$. Either *Pnma* or *Pbnm* is used as the subscript in these cases.

### D. Multislice simulation of 4DSTEM datasets and matching to experiment

Simulated 4DSTEM datasets were calculated from the structure of Bull *et al*. [24] using a multislice approach with the Dr Probe package [26], for a sample thickness of 53 nm. For the multislice simulations of the structure in $[111]_M$ orientation, a supercell of size $a_{SC}$=3.84731 nm, $b_{SC}$=3.89183 nm, & $c_{SC}$=1.10062 nm and orthogonalised unit vectors was set up and partitioned into 8 slices along the $\mathbf{c}_{SC}$ direction, i.e. the direction parallel to that of the incident electron probe. The slices were set such that each contains one atomic plane. Projected scattering potentials were prepared on a grid of 800 × 864 pixels for the ($\mathbf{a}_{SC}$,$\mathbf{b}_{SC}$) planes. For simulating the effects of thermal diffuse scattering within the quantum excitation of phonons (QEP) model [27], 100 atomic configurations were prepared for each slice. High angle diffraction patterns were calculated for a pixelated detector of 384 × 384 pixels placed in the diffraction plane (far field) with a pixel step of 0.25 mrad, assuming the experimental STEM parameters noted above and for every point in a scan at a similar pixel sampling to the experimental data.

This simulated 4DSTEM dataset was processed in the same way as the experimental data, and when comparing to experimental datasets averaged from many unit cells, it should be noted that the theoretical simulation is also the average of 100 different QEP simulations.

The experimental dataset for comparison was interpolated by a factor of 4 using *scipy.ndimage.zoom* and then 121 unit cell origins were found on La-O columns using *Atomap* [28] and used to determine image patches for superposition and summation – the interpolation meant that the results were not smeared by the origin positions varying at the subpixel level.

### E. Mapping of the displacements in 3D

To convert a map of fit parameters to a map of **b**-axis orientations firstly needs sets of peak positions to be determined with *Atomap* [28]: La-column HAADF peaks, and the peak positions in the $A_1$ map (being easier to quantify reliably than the trough positions). These are then sorted by coordinates and nearest-neighbour $A_1$-site peaks are associated with each La column peak (i.e. HAADF peak) in sorted lists. These are then used to and give the sense of the **b-**axis in 3D (not just projected into 2D). The angles are determined from the average of a patch around each HAADF peak (3x3 in this work, but this would be easy to adjust to 5x5 if desired). The strength of the La position modulation is determined from the $A_2$ values at the peaks of $A_1$ (the troughs are more difficult to find reliably). The resulting data can then be plotted as a map of **d** orientation, and modulation strength (atom off-centre shift), normalised to the maximum value observed previously by structure refinement [24]. This plot can be then displayed using a range of representations (e.g. arrow or "quiver" plot, colormap). Note, whilst we expect the **d**-direction to be parallel to **b**, it is plausible that it could deviate from such a crystallographic alignment locally, especially close to boundaries or defects, so the orientation is worth mapping at high resolution.

## III. RESULTS

### A. Fitting the azimuthal diffracted intensity variation at atomic resolution

We showed recently that unidirectional atom position modulations as shown in Figure 1 lead to a strong periodic azimuthal intensity variation in the First-Order Laue Zone (FOLZ) ring clearly aligned with the modulation direction. It was shown that this can be used to determine the modulation direction and strength through fitting the azimuthal variation to a simple periodic function [29], in a way that is far more accurate and informative than simple atomic resolution imaging. However, this was just performed on larger image patches covering several unit cells and no attempt was made to do so at atomic resolution, or to investigate the details of what might emerge from such fitting at atomic resolution. **Figure 2**b) shows a representative diffraction pattern averaged over a significant image area (to improve signal to noise) with the FOLZ indicated within dotted lines describing two concentric circles. The data was polar transformed and then be integrated to a single intensity plot as a function of azimuthal angle (measured anticlockwise from horizontal-right) for fitting to Equation (1). The fit shown in Figure 2a clearly peaks close to 90° and 270°, with additional intensity on the 90° side (top of diffraction pattern). The remainder of Figure 2 shows the results of applying this approach to every pixel of an atomic resolution 4DSTEM scan of a single crystal area of a $La_2CoMnO_6$ thin film (grown on $LaAlO_3$-$(Sr,Ta)AlO_3$ - LSAT)[25]. A HAADF image was calculated from the polar transformed data by simply integrating all intensity above a scattering angle of 79 mrad and this shows the expected appearance for a perovskite viewed along a $\langle 110 \rangle_C$ direction. The centres of the brightest peaks in the upper half of the image (as determined using Gaussian interpolation), arising from La-O columns, are indicated by fuchsia stars, and these stars are overlaid on the equivalent images from the FOLZ fitting. Figure 2d-g) show the best fitting $A_2$, $\phi_2$, $A_1$ and $\phi_1$ parameters of equation (1) where $A_2$ is the amplitude of the 2-fold intensity modulation, $\phi_2$ is the peak angle of that 2-fold modulation (defined in the range 0-180°), $A_1$ is the amplitude of the 1-fold intensity modulation, and $\phi_1$ is the angular shift of that 1-fold modulation (defined in the range 0-360°). $B$ is not shown as it is just the offset of the base level of the FOLZ from zero, which is mainly a function of specimen thickness/density and is rather similar in form to the HAADF image.

The first overarching point is that all these parameter maps are very consistent and show clear repeating patterns across each unit cell which do not simply mirror the HAADF intensity and clearly contain information absent from the projection-only HAADF image. Almost all prior atomic-resolution imaging techniques in scanning transmission electron microscopy have shown features that are either related directly to the projected potential (annular dark field [30,31], bright field [8], annular bright field [32], ptychography [33,34], iDPC imaging [35]) or the derivative of the projected potential (atomic resolution differential phase contrast [17] and first moment or centre-of-mass imaging [36,37]). The only recent departure from this in atomic-resolution imaging has been *Symmetry-STEM* (S-STEM) in which 4DSTEM diffraction patterns are assessed by applying a symmetry element (e.g. a rotation) and cross-correlating with the original and plotting the cross-correlation coefficients as a function of position [38]. In that case, unprecedented detail is seen that is not simply related to projected potentials, and the details have only partially been explored, however the method has not yet been widely used at atomic resolution. Also, it only uses small diffracted angles in the First Order Laue Zone (basically just low order diffraction spots overlapping with the primary beam). Thus, mapping features related to high angle electron diffraction to a Higher Order Laue Zone, whatever the precise interpretation thereof, reveals things never previously seen in atomic resolution STEM imaging. In this work, the details of what is seen in this specific case are described first, prior to investigating what physically meaningful information can be determined from these maps by comparison to theory.

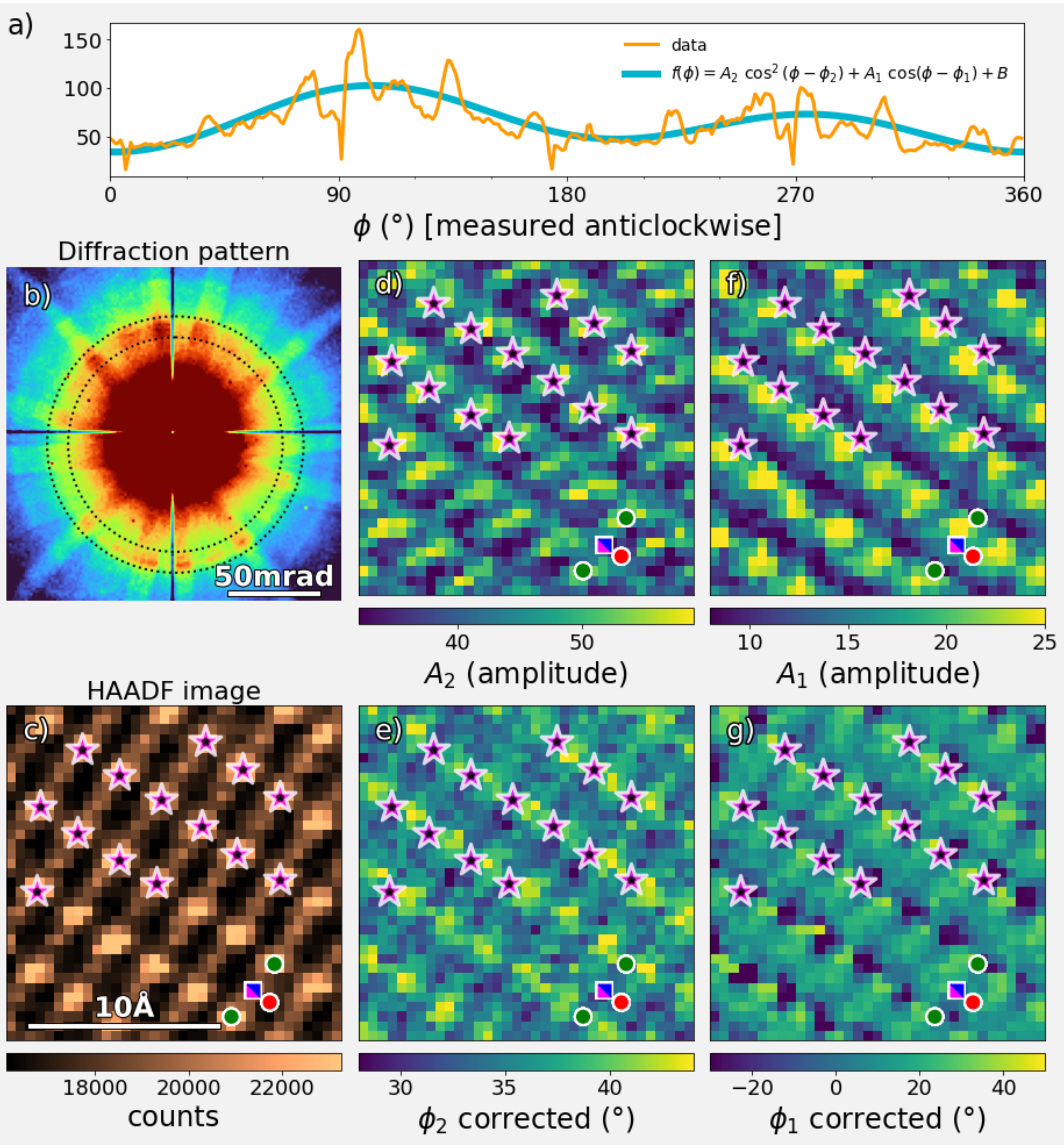

FIG 2. Atomic resolution fitting of the First Order Laue Zone ring in a region of a $La_2CoMnO_6$ film calculated from a 4DSTEM dataset recorded along a $\langle 111 \rangle_M$ direction. This shows b) a sum diffraction pattern from the whole area with the FOLZ ring indicated with dotted lines; a) the fit of the ring from this sum pattern after polar transformation; c) a HAADF image calculated from the dataset for angles >79 mrad; and d), e), f) and g) fit maps for the parameters $A_2$, $\phi_2$, $A_1$ and $\phi_1$ as defined in equation 1 and the legend of part b). An overlay of La-O atomic column positions is given on maps c)-g), together with a unit cell key (green – La-O, magenta/blue – manganese/cobalt, red – O).

For specific parameters, the $A_2$ map shows a hitherto unexpected feature where the strength of the 2-fold oscillation does not peak at the position of the atom columns in the HAADF image but appears elongated either side of the atom column centre, and is small elsewhere – with a much wider elongation of ~ 1.3 Å that the projected column distortion of about 0.2 Å. The $\phi_2$ map (corrected in angle as clockwise from $[1\bar{1}0]$, assuming the beam direction is [111]) is centred at around 35°, as expected [29]. Nevertheless, this angle twists near the La-O column positions and around the $A_2$ peaks: to the lower right of each La-O column it peaks up to about 41-42° and in-between the La atom rows, it dips to below 35°. The $A_1$ map of the unidirectional intensity modulation is particularly surprising as it peaks away from the La column centres, towards one of the arms of the $A_2$-elongated peak. On the other side of the same La-O column, it shows a large dip in intensity. Finally, the peak angle for the unidirectional distortion measured by the $\phi_1$ parameter only shows small variations over most of the image but shows significant variations near the dips in the $A_1$ parameter (with very large uncertainties because it is poorly defined when the unidirectional modulation is weak).

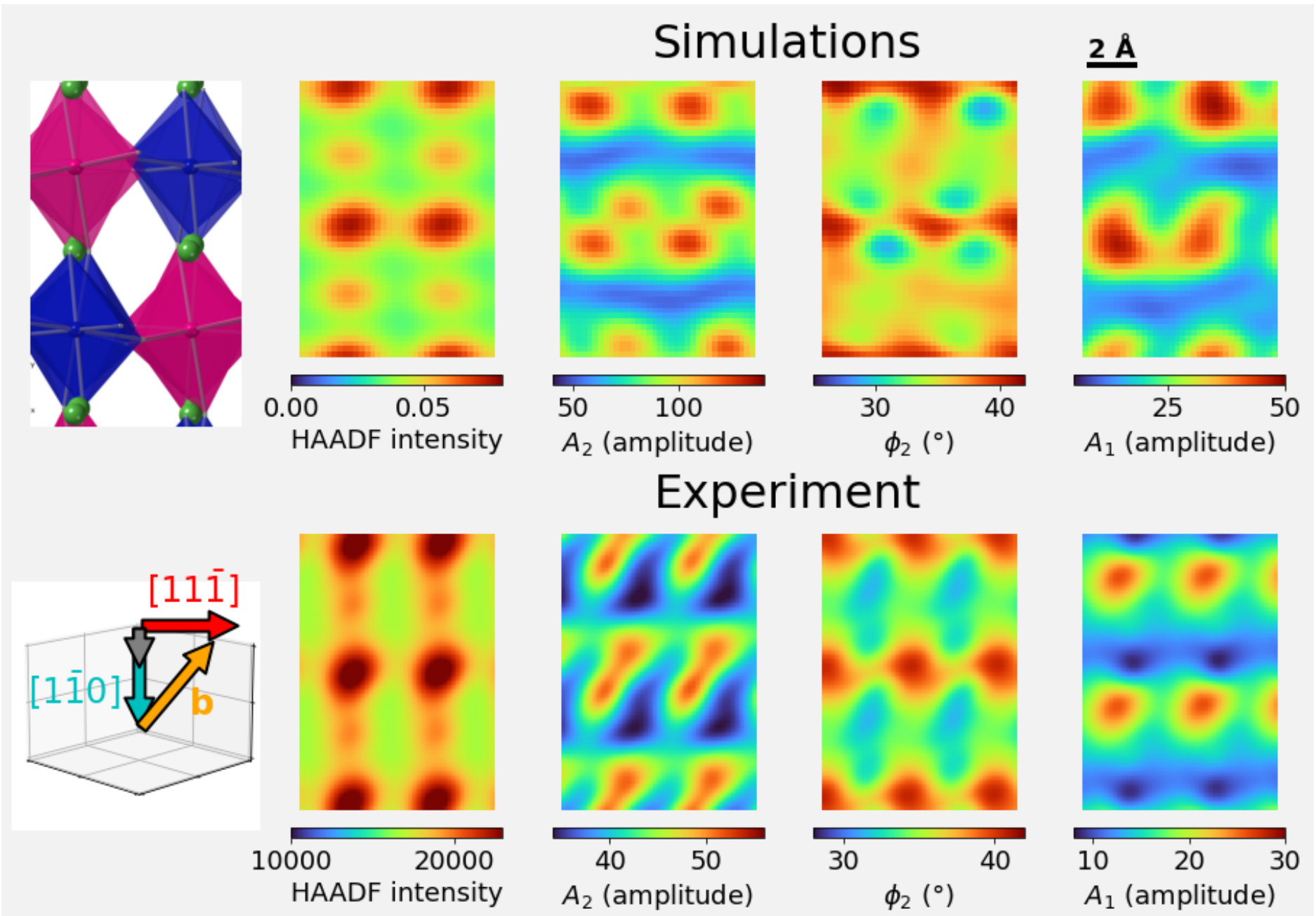

FIG 3. Comparison of azimuthal fitting applied to a simulated 4DSTEM dataset with a beam direction of $[111]_M$ (after upsampling by a factor of 2) compared to template averaging of experimental fits from the part of the same dataset as shown in Figure 2 (after upsampling by a factor of 4 before alignment to aid registration with ¼ pixel accuracy). Insets show a schematic diagram of the crystal structure in this projection (atom colours as for Figures 1 and 2) and a 3D plot of specific crystal directions in this figure (the orange arrow for **b** is pointing up, right and towards the reader), the grey arrow is the beam direction.

### B. Atomic resolution diffracted intensity shows the atom displacements

To interpret the atomic resolution detail in diffraction pattern fitting, multislice simulations [39,40] were performed for a suitable supercell of the structure followed by application of the same polar transformation and fitting as used on the experimental data using a $[111]_M$ direction. This is shown in the upper part of **Figure 3**, where HAADF intensity, and the $A_2$, $\phi_2$ and $A_1$ parameters are plotted (with an upsampling of a factor of 2 to make the plots a little smoother for easier interpretation). For ease of comparison with experimental data, a patch of the same experimental dataset as used to make Figure 2 was processed as before and the results then interpolated by a factor of 4 to allow subpixel alignment. Image patches around 121 different La-columns were then superimposed and averaged to make representative supercell images of each parameter. The visual comparison is evident and, whilst the simulations have slightly stronger contrast and better spatial resolution, it is clear that the theory and experiment are mostly in good agreement.

For a detailed explanation of the diffraction physical processes behind these patterns, please see the Supplemental Information section S3, but in short, two patterns should be borne in mind. Firstly, the 3D directions of the **b**-axis and the La-atom displacements makes the structure factors of reflections on opposing sides of the FOLZ ring unequal in magnitude. Secondly, electrons impinging on a crystal slightly offset from a heavy atom column tend to be pulled towards the column by Coulomb interactions and scattering is biased in this direction. These two effects combine to give much of the contrast seen in the maps.

The key practical point here is that a simulation using a refinement of the published bulk structure [24] produces a diffraction dataset that contains very similar features to those seen in the experimental dataset, which suggests that we can use the simulations for which we know the ground-truth as an interpretation key that allows us to determine crystallographic parameters from experimental maps. The first key useful thing to note is that the asymmetric dip and peak in $A_1$ magnitude to either side of the La-column centres is reproduced in theory for a perfectly aligned beam along the axis. This shows that the $A_1$ parameter is not merely due to mistilt, as we had erroneously assumed in our prior work [29]. It is also clear in this figure that the $A_2$ parameter map shows dumbbell-like structures peaked to either side of the column centre observed in the HAADF image (spacing about 1 3Å) which are like sharper versions the streaks or dumbbell like features in the experimental maps.

The calculation shown is for a $[111]_M$ beam direction (i.e. the direction pointing upwards, antiparallel to the direction of electron motion and parallel to the direction of current flow), with the [010] direction pointing to the upper right and up out of the page. This is in the direction from the $A_1$ peak to the $A_1$ dip for each La-O column, or alternately from the $A_1$ peak to the nearby HAADF peak. In view of the correlation between simulation and experiment, and the clear evidence in the Supplemental Information that all this is coming from interaction of the electron beam with the La-atom movements, this then provides an unambiguous method for determining the 3D orientation of the **d**-direction. (The reader is referred to the Supplemental Information Section S2 for a demonstration that the same applies for other $\langle 111\rangle_M/\langle 11\bar{1}\rangle_M$ type directions in this crystal structure). It is already known that the intensity of the FOLZ is approximately linearly dependent on the magnitude of atom position modulations along this **d**-direction [19].

Thus, by combining these the HAADF, $A_2$, $\phi_2$ and $A_1$ maps, we can get the i) strength of the modulations (from $A_2$), ii) the direction of the modulations (and therefore the direction of the **d**-direction with a 180° ambiguity in 3D) (from the $\phi_2$ parameter averaged from a few pixels (in this case 3x3) around each HAADF peak) and iii) the absolute sense of the direction in which the **d**-vector is pointing up out of the page (from where the peaks of the $A_1$ parameter lie with respect to the peak of the HAADF intensity).

## C. Mapping atomic displacements in 3D across an interface

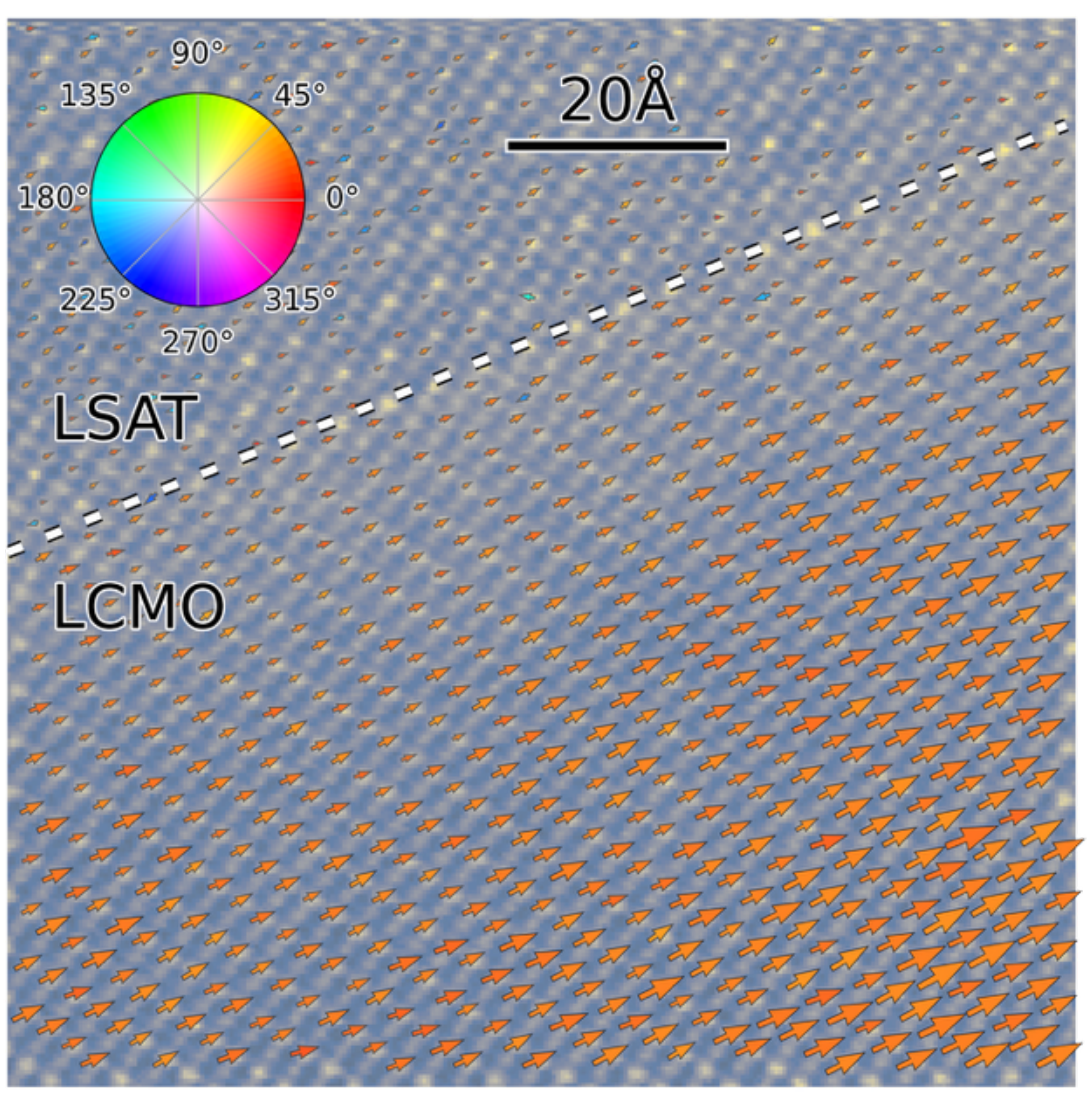

FIG 4. Atomic resolution imaging of 3D orientations of La atom displacements in a LCMO/LSAT interface area from a 4DSTEM dataset (LSAT substrate above the dotted line, LCMO below). **d** vector orientation and cation displacement are mapped with atomic resolution with arrows: the arrow direction show the direction in which **d** points out of the page, whereas arrow size shows $|\mathbf{d}|$, and colour correlates 3D direction in the upper hemisphere according to the colour wheel. This is overlaid on a HAADF image calculated from all electrons scattered above 79 mrad to show the correlation with a standard atomic-resolution projection image.

From the above, we have determined these parameters from a set of HAADF, $A_2$, $\phi_2$ and $A_1$

maps. In **Figure 4**, this is shown for a 4DSTEM scan across an area containing an interface between a LSAT substrate (top left) and a LCMO thin film. Using the procedure outlined above, Figure 4 shows arrows (aka a quiver plot), whose angles represent the directions of the **d-**vector for each atomic column, whose heads point in the direction in which the **d-**vector has a positive inclination (i.e. upwards from the plane of the page, so an up-right, down-left displacement pattern would give a right pointing arrow and an up-left, down-right displacement pattern would give a left pointing arrow), and whose length corresponds to the strength of the atom modulation along this direction. Additionally, the arrows are coloured according to the *HSV* colour wheel superimposed to represent the 3D direction, where *Hue* represents the azimuthal angle of the **d-**direction about the beam direction, and *Saturation* represents the component of the modulation in-plane (thus paler colours are more vertical), and *Value* is set to 1. In this case, since **b** directions have to be at 30° to the sample plane for any $\langle 111\rangle_M/\langle 11\bar{1}\rangle_M$ direction, so $S = \cos 30° = \sqrt{3}/2$. This map is superimposed on a semi-transparent version of the HAADF image as a guide to the eyes. The HAADF image shows the characteristic contrast of LSAT (above the interface) along $\langle 110\rangle_C$ where different patches show slightly different ordering because of internal chemical inhomogeneity. The LCMO (below the interface) shows the typical appearance for a simple perovskite along this beam direction (as also shown in Figure 2 for the HAADF intensity) with little obvious sign of atomic modulation.

The arrow map shows no clear modulation in the substrate, as expected, randomly oriented small values (average 2 pm magnitude with an average direction of 58° ± 73° [ACW from horizontal-right]). Throughout the LCMO, however, the modulations all point parallel to the interface towards the top right. The strength of this modulation is very weak for the first few unit cells above the interface (average 4 pm, 25° ± 22° in-plane angle in first 2 nm) but then gradually increases with distance from that interface (average 11 pm, 24° ± 2° in-plane angle > 6 nm from interface). This is in accordance with previous measurements from this very film [11,25], as well as many previous studies showing that strain and properties in thin perovskite films often increase to the full value in a few nm from the substrate, as a result of the constraining effects of the interface [18,19,41,42].

Previous work on LCMO growth concentrated on the effect of using a $\{111\}_C$ growth surface to promote the octahedral tilting necessary to create a periodic modulation of B-site size which then results in ordering of Co and Mn, and thereby strong antiferrimagnetic ordering with a strong magnetic moment [25]. However, the present result indicates that this growth surface promotes in-plane **b**-axis orientation of the crystal structure, bringing the A-site modulations into the substrate plane. It was speculated in our prior work [25], based on strain considerations, that this might be the case for films on LSAT, but no evidence was presented at the time to confirm or deny this hypothesis. The same trend of the the **b**-axis lying in plane has been seen in seven atomic-resolution datasets from different areas of the same film. A similar approach has also been applied to a dataset from the film recorded over a much larger image area (not at atomic resolution) and the same still holds true (although absolute mapping of the sense of the **b**-axis inclination is then not possible because of not having both the atomic resolution imaging of column positions and the $A_1$ parameter peak positions); this is shown in the Supplemental Information (Figure S7). As has previously been concluded, octahedral tilting and A-site modulation are all connected in complex oxides like this [19,25]. A key outcome of the present work is that promoting a particular A-site modulation direction appears to be a powerful way to control the octahedral tilting.

## IV. DISCUSSION

### A. Atomic resolution 3D displacement mapping gives new insights into thin film growth

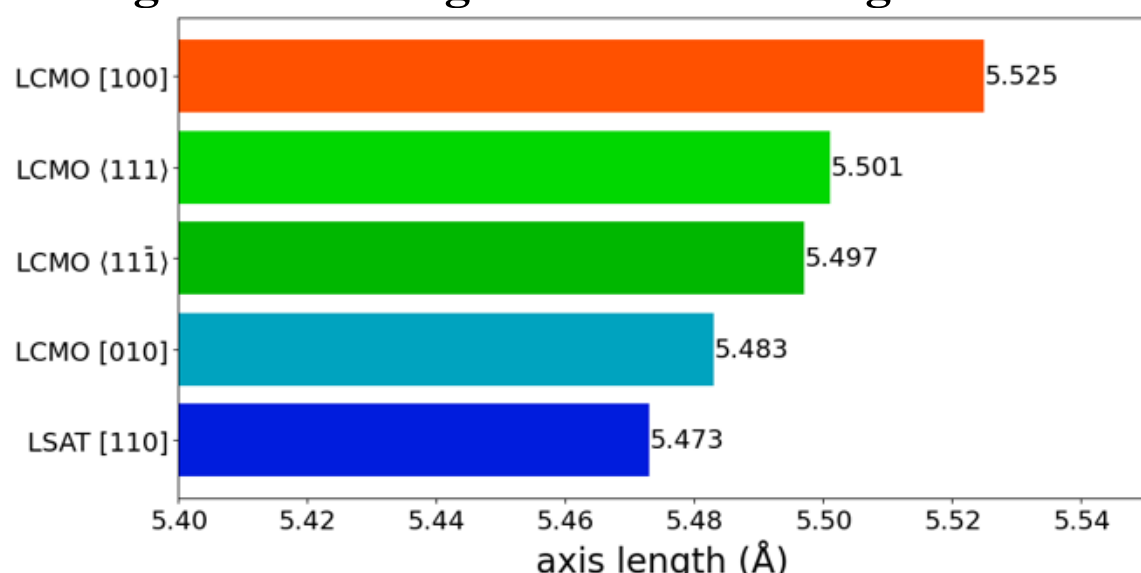

FIG 5. Axis lengths for different directions in LCMO equivalent to cubic ⟨110⟩ directions, and a comparison to the [110] length in the LSAT substrate.

This brings us to the question of why the A-site modulation should be preferentially in-plane. Generally, domain orientations in perovskite thin films have mainly been found to be determined by strain minimisation (for instance Lebedev *et al.* [43], MacLaren *et al*. [44]). As it turns out, this may be also true here. This is summarised visually in **Figure 5**. LSAT should have a relatively short pseudocubic ⟨110⟩ vector of length 5.473 Å. The **b**-axis of LCMO should be 5.483 Å in length (so 0.18% compressive strain), $[11\bar{1}]_M/[1\bar{1}\bar{1}]_M$ should be 5.497 Å (0.46% compressive strain) and $[111]_M/[1\bar{1}1]_M$ should be 5.501 Å (0.52% compressive strain). So, all of these could be relatively low strain if laying in-plane. In contrast, the **a**-axis of LCMO should be 5.525 Å, which is a significantly higher strain of 0.94%. So, the most energetically favourable configurations will be with **b** and two $\langle 111\rangle_M/\langle 11\bar{1}\rangle_M$ axes in plane on the (111) surface of the LSAT (in other words, a $(101)_M$ film plane [45,46]). This results in a slightly plane larger spacing normal to the interface (2.25 Å unstrained) than it would for **a**-axis in-plane (2.24 Å unstrained) (ignoring any Poisson's ratio effects which would probably increase this further [41]), which probably creates more space to incorporate ordered layers of larger Co atoms. Also, Kjaernes *et al.* [46] reported that $(101)_{Pbnm}$ growth promotes a strong modulation of $(101)_{Pbnm}$ plane spacings, giving alternate larger and smaller plane spacings out-of-plane, which would create larger and smaller spaces for B-site atoms, among other things. This may well be the reason for better Mn/Co ordering on (111) LSAT than growing on (111) $SrTiO_3$ [25] (which has a larger lattice parameter that would probably not favour the same domain orientations). And better Mn/Co ordering gives better ferrimagnetic ordering and a stronger magnetic moment, and thus better properties. This work therefore provides a method for understanding the structural origins of these improved properties and imaging in atomic-resolution detail how those growth choices are affecting the structures that give rise to the important functional properties.

One thing that is clear is that this level of detail about domain structures and how they relate to distinct crystallographic axes in characterisation techniques was not seen in any characterisation techniques used previously. Perhaps with a lot of sample tilting in the TEM goniometer, evidence could be found for some domains with **b**-directions in plane using TEM nanobeam diffraction (this would require getting [010] on-axis, and since this would require tilts of >60° for some domains, is not feasible for most domains). It is possible that the differences between **b**-axis in-plane and **a**-axis in-plane could be resolved by careful thin film X-ray diffraction. But this method of fitting the high angle electron scattering provides much more straightforward and highly spatially resolved information than any prior method.

### 4.2. Wider significance of this work – atomic resolution crystallography using STEM

In the specific case of perovskite oxides with the $GdFeO_3$ structure (to which the double perovskite structure examined in this work is related), such structures involve an orthorhombic unit cell (or monoclinic or triclinic distortion thereof) with a $\sqrt{2}a \times \sqrt{2}a \times 2a$ cell size and a modulation of the A site atoms along a direction close to the crystallographic **b**-axis. The HOLZ-STEM method was originally developed on one of these – $LaFeO_3$ [19] on $SrTiO_3$. However, these are fairly commonly used in functional oxide heterostructures.

One frequently used example is $DyScO_3$, especially for perovskite substrates with a relatively large lattice parameter (pseudocubic a = 3.97Å). However, this has a big difference in a and b parameters of about 5%, and a large atomic displacement of these La atoms. Clearly, this is going to affect the structure of the interface between it as a substrate and any film grown epitaxially upon it. Boese *et al.* [47] found that $DyScO_3$ grown in a heterostructure with (100) $SrTiO_3$ had either $[110]_{Pnma}$ or $[001]_{Pnma}$ in-plane. This was performed by careful inspection of contrast in High-resolution TEM. Also using the apparent interface plane in the images, this means that $[1\bar{1}0]_{Pnma}$ is the out-of-plane direction. This was found subsequently to lead to a puckered interface and a reconstruction where every other interface atom column of Dy is much brighter [48] when viewed along $[110]_{Pnma}$. Now this is relevant way beyond just being a structural curiosity, since these $DyScO_3/SrTiO_3$ interfaces show high conductivity under certain circumstances [48,49], and such structural rearrangements at the interface dependent on 3D crystallography may well play into such properties. The method shown in the present work provides a far more sensitive and detailed characterisation of this 3D crystallography than would have been possible previously, including showing a capability to quantify any change of A-site modulation strength or direction close to the interface.

As another example, $SrRuO_3$ is also a perovskite with the $GdFeO_3$ structure, but is intriguing as it is a ferromagnetic metal. As such, it has been widely used as an epitaxially lattice-matched contact in oxide heterostructures as well as for the magnetic properties. It also can show fascinating interfacial properties, such as a topological Hall effect in interfaces with $SrIrO_3$ [50] and interesting charge transfer leading to conductivity in the $SrIrO_3$ [51]. However, in those studies, no consideration was given to any effects of orthorhombic crystal orientation nor was this looked for. However, it can reasonably be expected that the band structure will be directionally dependent and that growth to favour different surfaces could have a significant influence on growth. The method shown in the present work could provide a lot more information about how the atomic modulations and crystal axes behave in practice if such a study were undertaken.

For a third example, growth of $LaFeO_3$ has been shown to be different on different $(111)_C$ surfaces of different orthorhombic perovskite substrates for growth on $(101)_O$ or $(011)_O$ facets [45] (using $DyScO_3$, $NdGaO_3$, and Gd $ScO_3$). That was important for magnetic properties, since this allowed controllable switching of the magnetic easy axis. Additionally, there are hints in the RHEED data that the crystallographic distortions are reduced close to the interface, possibly similar to the reduced displacements close to the substrate interface in the present work. However, that work contained no microscopy and either $(101)_O$ or $(011)_O$ growth would leave a possibility of domain formation with different **b**-axis orientations. How these different domains contribute to micromagnetic textures was unexplored. Combining the methods provided in this work with recent advances in Lorentz imaging of magnetic induction [52] would be a promising way forward for understanding substrate/interface control of magnetic properties in a film.

More generally beyond perovskites, this work shows a way forward for mapping key parameters from 3D crystallography with atomic column resolution. Any feature of a diffraction pattern that varies across a unit cell as a function of radial or azimuthal angle could be extracted by similar approaches to these of fitting data with appropriate mathematical functions, and if applied to 4DSTEM data taken at suitable spatial resolution, could be investigated at the atomic scale. For example, any modulation or chiral displacement of atoms along a column should leave interpretable signals that could be determined and mapped from intensity distributions on a HOLZ ring. Of course, each case would need validation by simulation to enable the interpretation of the atomic resolution diffraction signals, in a similar way to in this work. This will be especially powerful in observing the variation of crystallography at the nano- or atomic-scale (i.e. basically inaccessible to traditional bulk diffraction techniques) in the vicinity of internal interfaces and defects in materials and heterostructures. It would also be incredibly powerful at elucidating the details of structure and composition on a column resolved scale in larger unit cell complex oxides and other complex compounds, where the structure solution from single crystal X-ray or neutron methods is ambiguous and somewhat depends on starting guesses for the occupancy of different sites (e.g. the work of Azough *et al*. [23]).

## V. CONCLUSIONS

In summary, we show that it is possible to directly map the azimuthal variation in intensity of a First Order Laue Zone ring in electron diffraction of a double perovskite in atomic resolution 4DSTEM imaging. This can be simply parametrised as the amplitudes and peak angles for 2-fold and 1-fold trigonometric functions of azimuthal angle. The resulting parameter maps reveal a wealth of atomic resolution detail when applied to imaging of a $La_2CoMnO_6$ (LCMO) film along a $\langle 111\rangle_{monoclinic}/\langle 110\rangle_{cubic}$ direction. With the aid of theoretical simulations, this allows the unambiguous mapping of atomic displacements in three dimensions from a single scan. This showed the expected suppression of displacement close to the substrate interface, but a rapid rise to close to the bulk value in about 6 nm. However, this also showed a novel insight, specifically that the **b**-axis sits preferentially in the plane of the interface. Further analysis showed that this likely occurred for strain minimisation reasons, but would have been hard to see by techniques available prior to this work (including standard techniques in STEM imaging). More generally, this work demonstrates a new approach to atomic resolution imaging in the electron microscope that is not simply projection imaging, but explicit works on mapping of 3D crystallographic parameters in atomic resolution images from fitting diffraction features in atomic resolution 4DSTEM datasets. It is anticipated that this broad concept will have a major impact on the atomic scale understanding of materials and nanostructures in the future, especially in systems where the crystallography is all tied to a single crystal orientation of some fundamental lattice such as epitaxial systems, domain structures, displacive transformations, and order-disorder transitions.

## DATA AND CODE

Will be available for download once this is past review since that needs a library archive to be set up…

## ACKNOWLEDGEMENTS

This work would never have been possible without the EPSRC funding of the grant “Fast Pixel Detectors: a paradigm shift in STEM imaging” (EP/M009963/1), which started the work that led to this and funded the computer server used in the calculations. We thank Diamond Light Source for access and support in use of the electron Physical Science Imaging Centre (Instrument E02 and proposal number EM16952) that contributed to the results presented here. The films were grown with the support of the European Research Council (ERC) (Advanced Investigator grant ERC-2009-AdG-247276-NOVOX), the EPSRC (Equipment Account Grant EP/K035282/1) and the Isaac Newton Trust (Minute 13.38(k)). IM is grateful to Prof Peter D. Nellist for helpful discussions as part of the EPSRC grant named above and to the late Prof John C.H. Spence for helpful discussions and encouragement to pursue work on HOLZ in 4DSTEM.

# Atomic resolution imaging of 3D crystallography in the scanning transmission electron microscope – Supplemental Materials

Ian MacLaren[1], Aurys Silinga[1], Juri Barthel[2], Josee Kleibeuker[3,4], Judith L MacManus-Driscoll[3], Christopher S Allen[5,6], Angus I Kirkland[5,6]

1. School of Physics and Astronomy, University of Glasgow, Glasgow G12 8QQ, UK
2. Ernst Ruska-Centre (ER-C 2), Forschungszentrum Jülich GmbH, 52425 Jülich, Germany
3. Department of Materials Science & Metallurgy, University of Cambridge, 27 Charles Babbage Road, Cambridge CB3 OFS, UK
4. DEMCON high tech systems Eindhoven BV, Kanaaldijk 29, 5683 CR Best, The Netherlands
5. electron Physical Science Imaging Centre, Diamond Light Source Ltd., OX11 0DE, UK
6. Department of Materials, University of Oxford, Parks Road, Oxford OX1 3PH, UK

## S1. Crystallographic proof of the displacement direction

The crystallographic unit cell of LCMO has a *P2₁/n* symmetry (Space Group #14). The cell has a multiplicity of 4 so atoms inserted at a general location produce the following 4 crystallographically equivalent Wycoff positions:

$$p_1 = x, y,\ z];\ p_2 = \frac{1}{2} - x, \frac{1}{2} + y, \frac{1}{2} - z;\ p_3 = -x, -y, -z;\ p_4 = \frac{1}{2} + x, \frac{1}{2} - y, \frac{1}{2} + z \qquad \text{[S1]}$$

Now for structures based on the $NdGaO_3$ structure, like LCMO, $x$ and $y$ are small and $z$ is close to 0.25 (actually $[0.0057, 0.0216, 0.2438]$ for LCMO [1]) and it is trivial to show that if we project half a cell along [111], then $p_1$ and $p_4$ line up with a small displacement:

$$p_4 - \left[\frac{1}{2}\frac{1}{2}\frac{1}{2}\right] = [x\bar{y}z] \qquad \text{[S2]}$$

Thus, effective displacement is given by

$$\mathbf{d} = \frac{p_1 - p_4 + \left[\frac{111}{222}\right]}{2} = [0y0] \qquad \text{[S2]}$$

The same calculation also works with $p_2$ and $p_3$. Moreover, if we project along any other $\langle 111\rangle_M$ or $\langle 11\bar{1}\rangle_M$ direction, this still works, even if we have to put a shift of a whole integer lattice vector in to overlap the atoms (effectively from neighbouring cells). For instance, along $[1\bar{1}\bar{1}]$:

$$\frac{p_1 - p_4 + \left[\frac{1}{2}\frac{-1}{2}\frac{-1}{2}\right] + [011]}{2} = [0y0]$$

Thus, the alternate displacements of the atoms along the $\pm\mathbf{b}$ directions are what we are seeing when viewing along $\langle 111\rangle_M$ or $\langle 11\bar{1}\rangle_M$, whether this is in atomic resolution STEM imaging, leading to elliptical columns, or in the present HOLZ-STEM method, leading to the azimuthal variation in intensity mapped in at atomic resolution.

S2. Demonstration that this mapping works consistently for $\langle 111\rangle_{monoclinic}$ directions

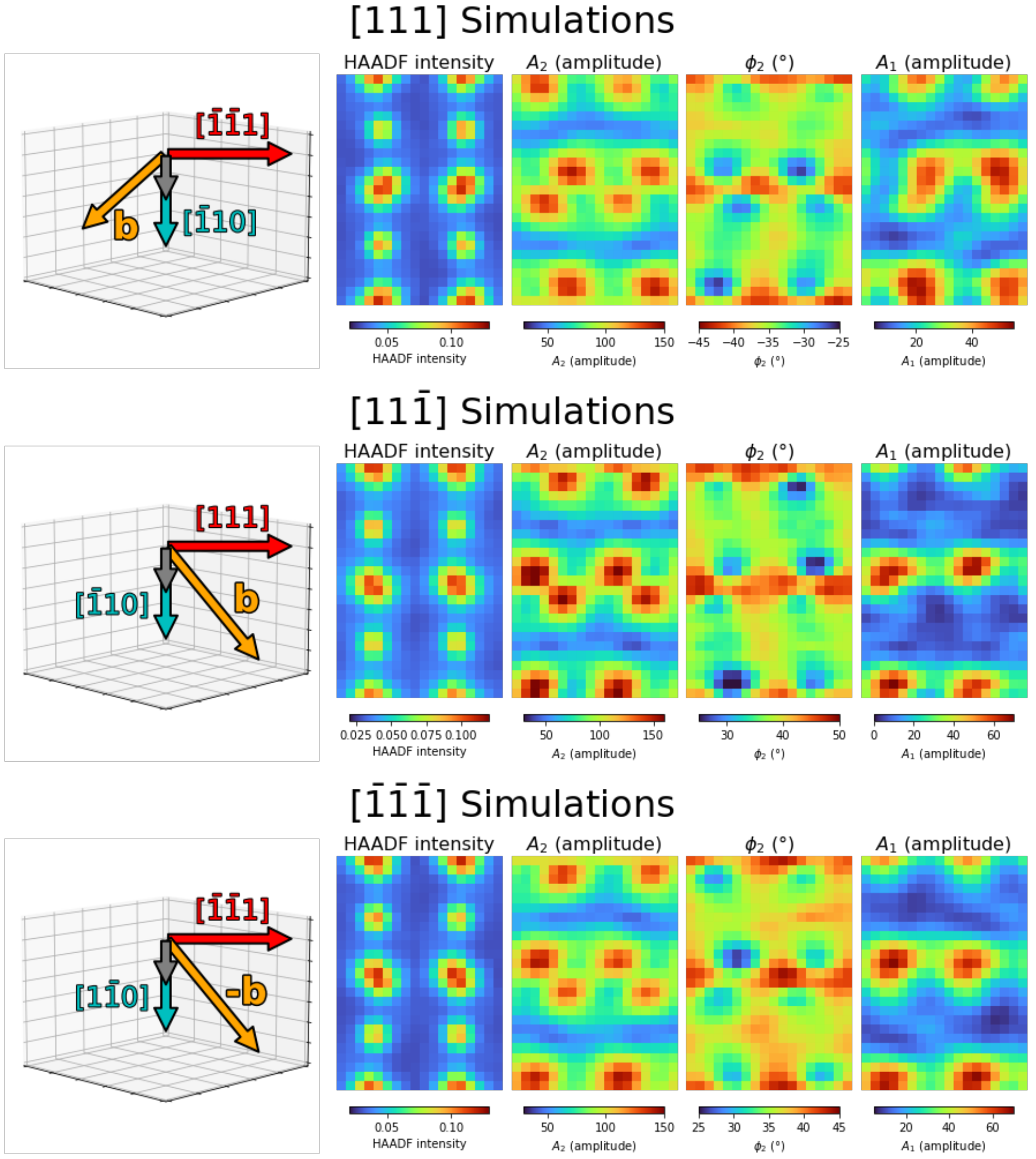

*Figs S1-S3: Fitting results for 4DSTEM simulations for three beam directions representative of all $\langle 111\rangle_{\mathrm{M}}$ directions. Detailed commentary given in the text.*

Simulated 4DSTEM datasets were calculated for several different $\langle 111\rangle_M$ directions corresponding to the main distinct ones in this monoclinic structure. It is in a standard setting with $\beta \neq 90°$, thus, there is basically no difference between $[uvw]$ and $[u\bar{v}w]$. There is, however, a small, but significant, difference between $[uvw]$ and $[uv\bar{w}]$ because the **a** and **c** axes are not orthogonal. Thus, $[111]_M$ and $[\bar{1}\bar{1}1]_M$ represent the two distinct types of $\langle 111\rangle_M$ directions. Including $[\bar{1}\bar{1}\bar{1}]_M$ demonstrates the effect looking along a direction in the opposite sense.

The first main point to notice is that very similar features come out in all maps. Note, the $\phi_2$ is measured anticlockwise from the vertical direction in all maps. This is reversed (as is the colour map) in Fig S1 where the "dumbbell" structure is pointing clockwise from vertical. So this parameter always tracks with the appearance of the maps. The really important point is that the direction in which the **b** axis (or -**b**) direction points up out of the sample plane is always in the direction from the $A_1$ peaks to the HAADF peaks.

A note on the 3D vector plots – these are done to roughly indicate the 3D direction of the vectors (although are done using a pseudocubic approximation of the axis lengths). To allow visibility of the grey Beam direction vector, these are tilted down by 11° so are not quite along the same direction as the Beam direction for the 4DSTEM datasets but are there to show the relative orientation of the key vectors in each image. In each case, the correct indices for the horizontal right direction (in monoclinic vectors) are given in red, and those for the vertical down direction in cyan. Beam directions are calculated with the convention as being vertical upwards from the page, antiparallel to the direction of electron motion and parallel to the direction of current flow.

## S3. A simple physical reasoning for the appearance of the patterns in the parameter plots

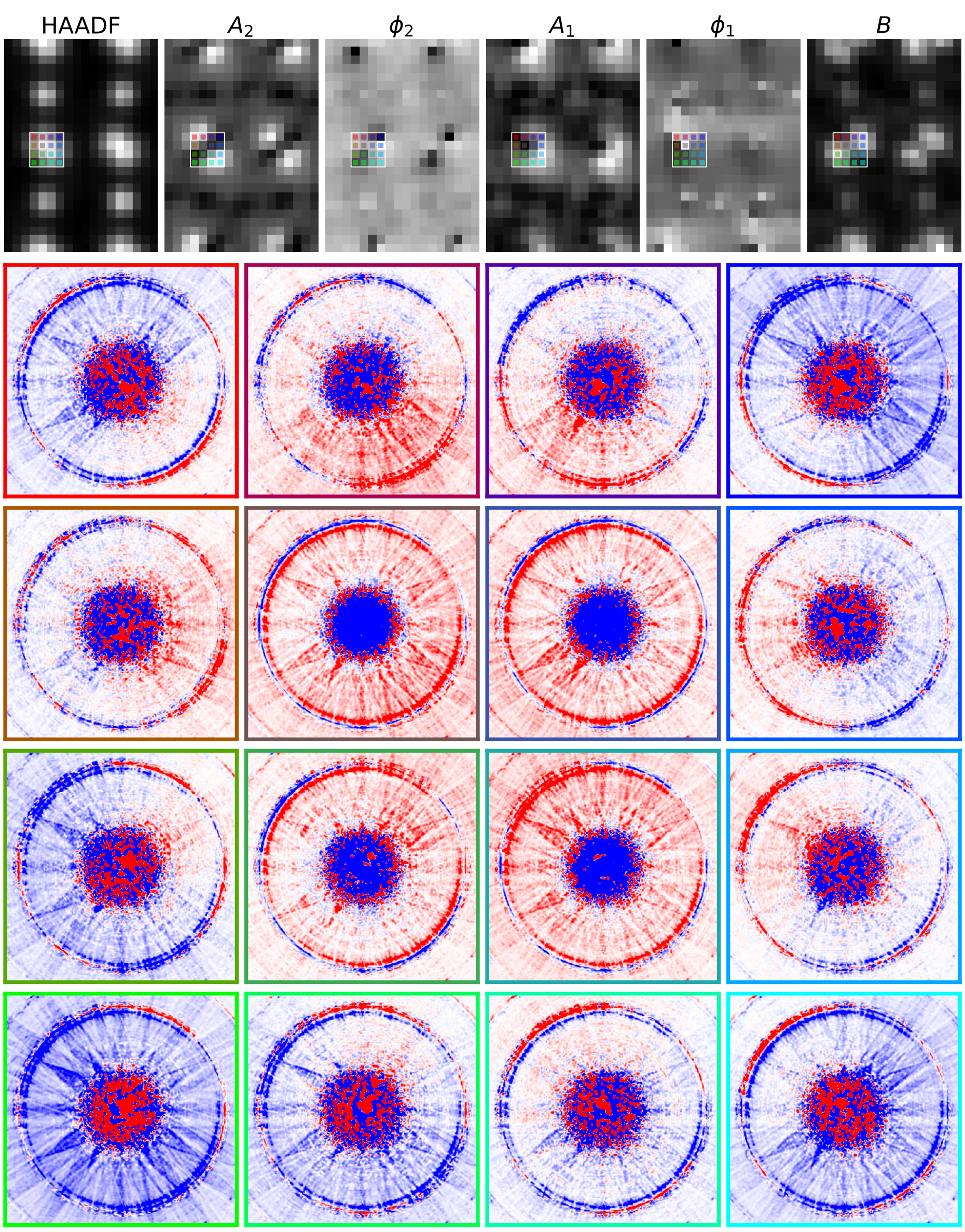

*Figure S4: Middle row shows parameter maps for a $[\overline{111}]$ simulation on a greyscale representation, with coloured points superimposed from which diffraction patterns were extracted and compared to the average pattern from the white box. Each pattern is displayed as the difference between the pattern from the individual point and the average pattern from the whole box over the La column, with red indicating excess intensity locally, and blue indicating less intensity locally.*

Firstly, in this structure, and all structures containing a symmetry breaking that results in a periodic modulation of a relatively heavy atom column, structure factors that would be equal in the primitive structure without La position modulation are no longer equal in magnitude. This, of course, was the basis of the original solution of such structures by X-ray diffraction. So, opposite positions on a HOLZ ring do not have to have equal structure factors. While dynamical diffraction is clearly strong and careful dynamical simulations (as used in this work) essential to a quantitative simulation, consideration of structure factors is still useful in the consideration of intensities in HOLZ rings as the extinction distances for such reflections are long and dynamical effects are much weaker than for low index spots close to the primary beam. Consequently, arguments based on kinematical scattering are already illustrative for HOLZ scattering in understanding the essential mechanisms.

To understand the details of how these basic asymmetries in FOLZ scattering are modified in an atomic-resolution 4DSTEM experiment, a set of simulations were performed using a $[\overline{111}]_{\mathrm{M}}$ beam direction, a supercell of $a_{\mathrm{SC}}$ = 4.39692 nm, $\mathrm{b}_{SC}$= 4.67019 nm and $\mathrm{c}_{SC}$ = 1.10062 nm, with potentials sampled on a grid of 800x864 points and a thickness of 53 nm, but otherwise as in the main paper. The simulation results were then examined by creating difference plots of diffraction patterns from a 4×4 grid of positions close to a modulated La column are shown in Figure S4, each showing the differences from the average pattern from the whole white box covering the column and its immediate neighbourhood. Each of the HAADF image and the 5 different parameter plots is shown with the grid overlaid on one La column to show where the diffraction patterns come from and there is also a colour coding of the overlay grid to the frame colour for each pattern.

The most obvious features of the whole set are that the central patterns, closer to the La column, are redder, showing that intensity has been pulled onto the column from the immediate surroundings by the normal Coulomb interaction. Meanwhile, the outer patterns are bluer overall, showing that they have lost intensity compared to the central column and are specifically not scattering so strongly to higher angles. However, looking closer, a second detail is visible in the outer patterns, which is that it is always redder at higher angles towards the column and bluer away from the column, always radially from the centre of the grid of patterns. Again, this is just the Coulomb attraction, and this is biasing any scattering to higher angles, including in the FOLZ, to directions towards the La column. This has specific and interesting effects for the formation of the FOLZ fitting parameter plots.

The first pair of interest are the red- and cyan-framed patterns taken from the two lobes of the $A_2$ plot in the top left and bottom right corners of the grid. In both cases, these are just outside the column and beyond the actual atom positions, but in the direction of the atom modulations. The principal feature in both cases is that they are close enough to the column to be getting a strong FOLZ coupling resulting in a strong $A_2$ signal. This coupling is exceptionally strong in these positions because the electron beam is interacting most strongly with *every other* atom in the column, since it is *every other* atom that is displaced towards this electron beam location (with the intervening ones displaced away from this beam location). So, the total intensity goes up in the FOLZ and its 2-fold modulation, but not anywhere else. However, the red-framed pattern is to the top left, and thus the attractive effects of the column will pull more intensity to the lower right side of the FOLZ (redder), and reduce intensity to the top left of the FOLZ (some blue seen). This fights against the intrinsic tendency from structure factors for upper left to be stronger, and thus the $A_1$ modulation is reduced. On the other hand, for the cyan-framed pattern from the lower right, the pull towards the column is in the same direction as the intrinsic asymmetry from structure factors and so the top left of the FOLZ strengthens, and lower right weakens, resulting in a stronger $A_1$.

Perpendicular to the direction of the modulation (blue and green frames), there is no strong coupling to either of the displaced sets of atoms on the column, but the arcs of intensity rotate a little around the FOLZ ring because of attraction to the column. Blue-framed, which is upper right, sees more FOLZ contrast pulled to the lower left, and less at the normal peaks at top left and lower right. Green, which is lower left, sees more FOLZ intensity at top right quadrant pulling towards the column and a reduction at lower right and upper left. In between these corners and the top-left and lower-right corners, there is some rotation of the peak intensity arcs on the FOLZ ring to follow the radial direction to the column centre, accounting for local rotations of $\phi_2$ in these areas.

## S4. Effects of convergence and defocus on atomic resolution mapping

Simulations were performed using the parameters also listed in section S2 on one supercell oriented along $[\overline{111}]$, varying the defocus from -20 nm to +20 nm, and this was performed for both the 20 mrad semiconvergence angle used in the experiment (Figure S5) and for 10 mrad semiconvergence (Figure S6). The results of then processing these simulated 4DSTEM datasets as before using cartesian – polar transformation, integration of the FOLZ ring, and fitting to the double cosine function are shown in Figures S5 and S6. As should be clear from a cursory investigation of these figures, a few trends are immediately obvious:

- If there is good atomic resolution contrast in HAADF, there will be good atomic resolution contrast in the fit parameter maps – the visibility of useful information in the two correlates extremely well
- In general, contrast decreases with increasing defocus in most maps, although a slight increase in $\mathrm{A}_1$ contrast is seen at -5 nm defocus, in general slight underfocus brings the probe to focus inside the sample and this gives some good contrast in the maps
- Defocus does not change the form of the $A_2$, $\phi_2$ or $A_1$ maps. The peak and trough to either side of the La columns in the $A_1$ maps never change places.
- The orientation of the extended feature about the La columns in the $\mathrm{B}$ maps flips orientation with defocus. This feature is not used in our algorithm for determination so, this one focus-dependent feature in the maps does not affect the work.
- Reducing the convergence angle to 10 mrad reduces the in-plane resolution but increases the depth of field over which the contrast is interpretable in terms of 3D orientations of La modulations

Drawing these points together, it is clear that as long as clear atomic-resolution HAADF contrast exists, the method is reliable and predictable with no complicating contrast reversals.

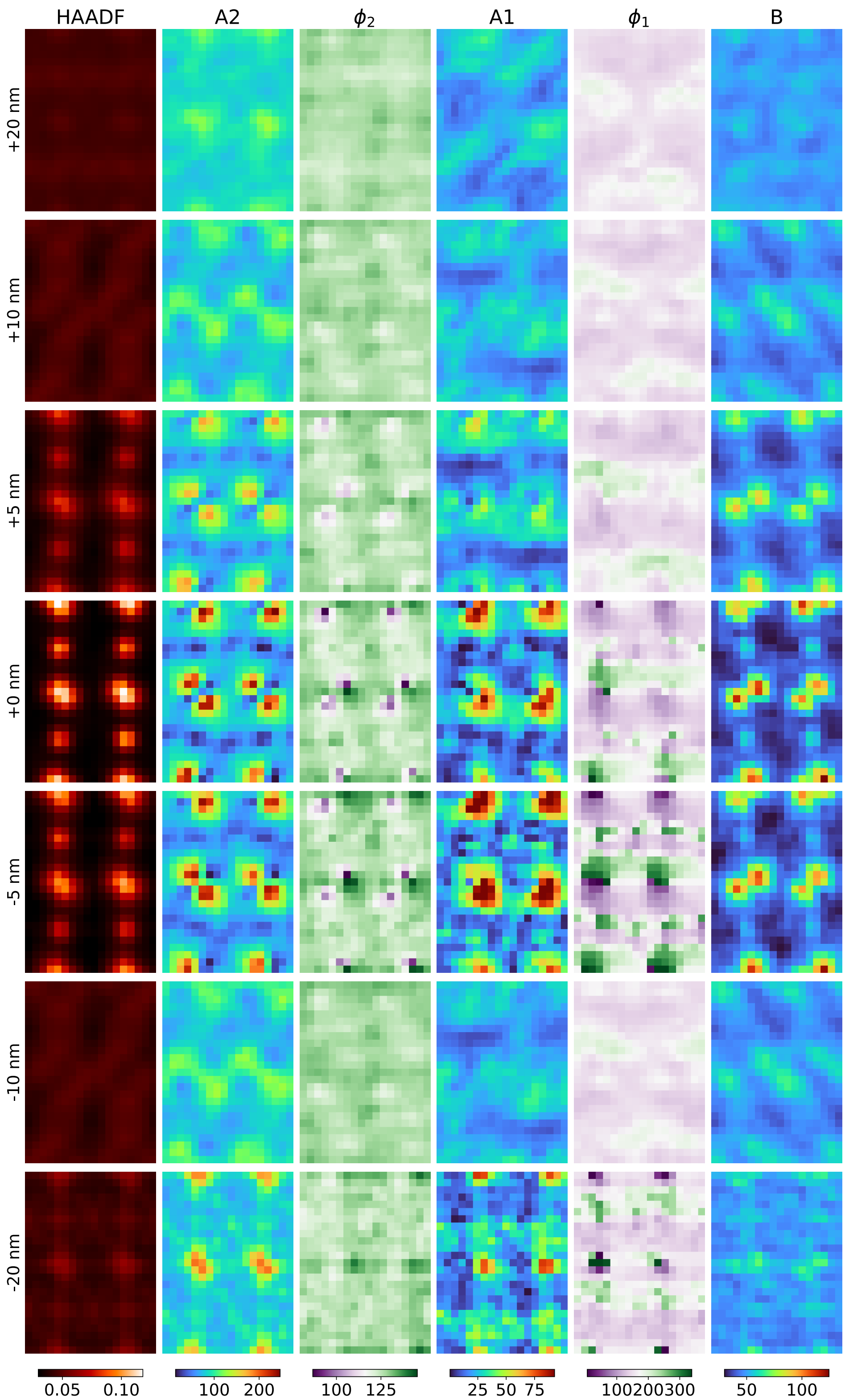

*Figure S5: The effect of defocus (from -20 - +20 nm) on the parameter mapping for the experimental choice of 20 mrad semiconvergence angle. All maps in each column are plotted on the same contrast scale represented by the colorbar at the bottom, in each case, the contrast limits are from the maximum and minimum on the map at +0 nm defocus.*

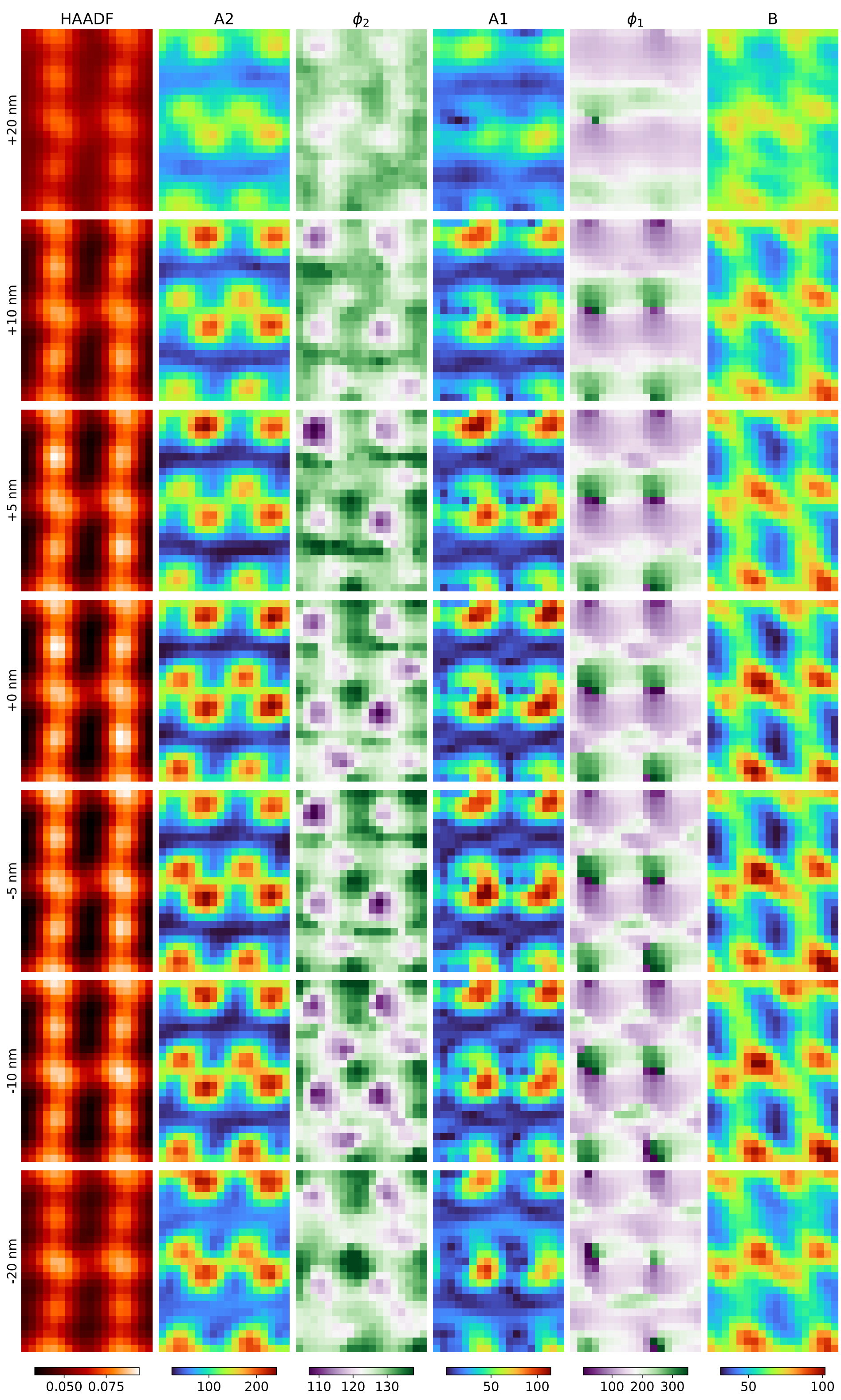

*Figure S6: The effect of defocus (from -20 - +20 nm) on the parameter mapping for a reduced semiconvergence angle of 10 mrad. All maps in each column are plotted on the same contrast scale represented by the colorbar at the bottom, in each case, the contrast limits are from the maximum and minimum on the map at +0 nm defocus.*

## S3. Large area mapping of same film (not at atomic resolution)

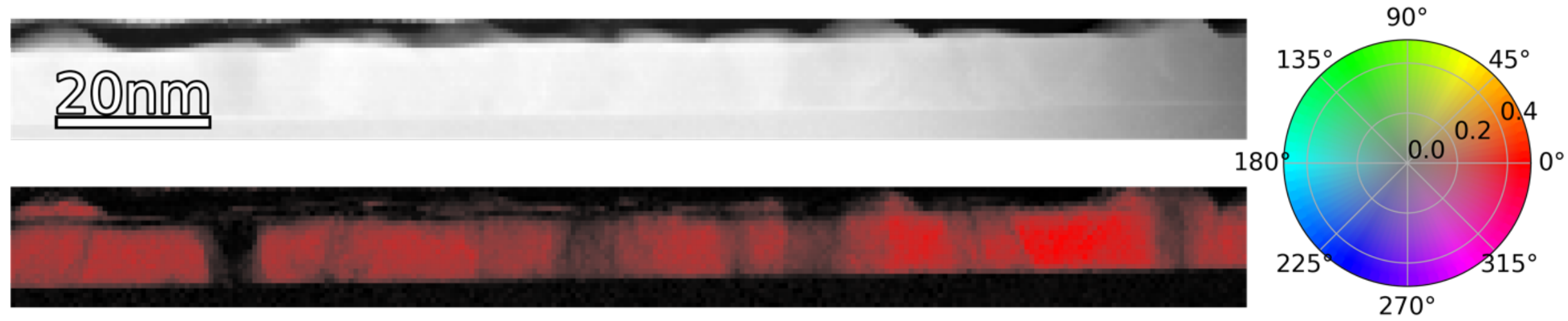

*Figure S7: Upper: HAADF image of the large area of this film with carbon to the top, film in the middle, and LSAT substrate at the bottom. Lower: large area mapping of the direction of the* **d***-axis for this film of LCMO on LSAT. All colour mapping of directions referenced to the image axes.* ***d*** *consistently lies close to in-plane suggesting that the* ***b*** *axis is in-plane for much of the film.*

Applying the method of this paper to a large area 4DSTEM dataset from the same film results in the following map in Figure S7. This shows that almost all areas have **d** laying in-plane at some angle to the specimen plane, suggesting the **b** lays preferentially in-plane. Some areas are dark in this map, which indicates regions where the inner FOLZ disappears and must correspond to other crystal orientations (separate evidence from low convergence angle nanobeam electron diffraction suggests these are [010] oriented domains). There is no evidence in the whole dataset for any area where the **b**-axis does not lie in the film plane.